# How good is your metalens? Experimental verification of metalens performance criterion

Jacob Engelberg,[†] Talia Wildes, [†] Chen Zhou,[‡] Noa Mazurski, [†] Jonathan Bar-David, [†] Anders Kristensen[‡], Uriel Levy, [†*]

[†] Department of Applied Physics, The Center for Nanoscience and Nanotechnology, The Hebrew University, Jerusalem, Israel, 91904

[‡] Department of Micro- and Nanotechnology, Technical University of Denmark, DK-2800 Kongens Lyngby, Denmark

KEYWORDS *Diffractive lens, Metalens, SNR, MTF, image processing, image simulation, imaging performance.*

**ABSTRACT:** A metric for evaluation of overall metalens performance is presented. It is applied to determination of optimal operating spectral range of a metalens, both theoretically and experimentally. This metric is quite general and can be applied to the design and evaluation of future metalenses, particularly achromatic metalenses.

## Introduction

Metalenses and diffractive lenses can allow miniaturization and economical mass production of optical systems by replacement of conventional lenses[1-3]. However, many applications require polychromatic operation (i.e. whenever the light source is not a laser), which seemingly cannot be supported by conventionally designed metalenses and diffractive lenses as a result of their strong chromatic aberration[4]. This drawback motivates recent research on the development of achromatic metalenses and diffractive lenses[5-7]. Unfortunately, the achromatization usually comes at the expense of reduced efficiency, lens power, and field-of-view (FOV). On the other hand, it has been shown that non-chromatically corrected metalenses, which we will call from here on 'chromatic metalenses', can be used over an extended spectral range, despite the performance degradation resulting from chromatic aberration[8,9]. It therefore becomes important to be able to compare overall performance of different types of metalenses, in order to find an optimal metalens design. In this paper we refer to metalenses for conciseness, but the results are equally applicable to diffractive lenses.

In order to evaluate overall metalens performance we must relate to both resolution and signal-to-noise ratio (SNR) and combine the two into a single performance metric. While many reports of metalenses include resolution and efficiency data, these performance metrics are not combined, so it is difficult to compare high-efficiency and low-resolution systems (characteristic of a chromatic metalens) to low-efficiency and high-resolution systems (characteristic of an achromatic metalens). Since degradation of resolution as a result of chromatic aberration can be compensated by a deconvolution image processing algorithm[10,11], at the expense of added noise, the efficiency and resolution metrics should not be separated. In a previous paper we proposed an ASNR (average signal-to-noise ratio) metric that fills this gap[12].

The purpose of this paper is to provide experimental verification of the ASNR metalens performance metric. As a case study, we apply this performance metric to determining the optimal operating spectral range for a chromatic metalens. However, the metric is more general, and can be applied to performance comparison of achromatic metalenses (dispersion engineered or spatially multiplexed) to equivalent chromatic designs – this will be the subject of a future paper. For a spatially multiplexed achromatic metalens - the method can be used to determine the optimal operating spectral range of each channel.

In our previous paper we theoretically described the ASNR metric and applied it to a generic metalens design[12]. In this paper we apply the theory to an actual metalens design and compare it to experimental results.

## Methods

Our measurements were performed on a wide-FOV metalens, with a focal length of f=3.36mm and F/2.5 (aperture diameter of D=1.35mm), operating around 800nm wavelength[13]. The measurement setup shown in Figure 1 consists of 3 parts: (1) Target projector (from light source to resolution target) (2) Camera with metalens (3) Spectral radiance meter (from L1 and on toward spectrometer and detector). The target projector provides a uniformly illuminated resolution target, which is imaged onto the camera by the metalens. The spectral radiance meter measures the absolute spectral radiance of the target (output end of fiber can be switched between spectrometer and detector). The alignment branch of the radiance meter (L3 and alignment camera) allows us to make sure the fiber input is "looking" at the correct area of the target. Additional details of the setup are given in section 1 of the Supporting Information.

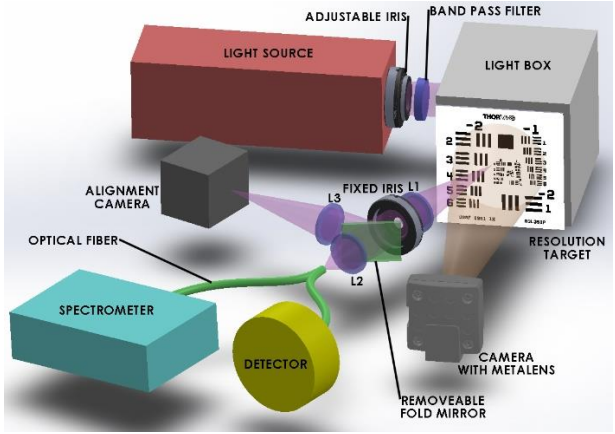

Figure 1. SNR measurement setup.

Our suggested metric for overall metalens performance is the ASNR, which is the SNR averaged over the relevant spatial frequencies of the image, i.e. from zero to the Nyquist frequency of the camera, as described by eq. 1 ($f_{nq} = 1/(2 \cdot pix)$, where $pix$ is the camera pixel pitch)[12]. The SNR that appears outside the integral is the zero (or low) frequency SNR. At higher frequencies the signal is attenuated by a factor equal to the modulation-transfer-function (MTF)[14] at that frequency, whereas the noise is the same for all frequencies (assuming a shot noise limited system, since shot-noise is generally white noise[15]. If this is not the case, the SNR can be placed inside the integral).

$$ASNR = SNR \int_0^{f_{nq}} MTF(v)dv \quad (1)$$

The ASNR can be measured directly or simulated based on system parameters. In the following we describe how both were done and compare measured to simulated results. This will allow us to experimentally verify the theoretical model. Once the model is verified experimentally, the proposed merit can be used to evaluate and optimize future metalens designs.

The measurement setup allows to measure the SNR directly, based on the video output from the camera (see section 5 of Supporting Information for characterization of the camera to show it is suitable for this measurement), and also to simulate the expected (shot-noise limited) SNR based on the spectral radiance. We performed the SNR measurements for several spectral widths using band-pass filters (BPFs).

The MTF of the metalens for the different spectral ranges was measured using a separate setup, described in [13], and was also simulated using Zemax optical design software. Comparison of measured to simulated MTFs has already been done in [13], but here we integrate the MTFs with the SNR to obtain the ASNR (per eq. 1), allowing us to compare simulated to measured overall performance.

The direct SNR measurement is performed by analyzing video images obtained from the metalens which is coupled to a video camera. We imaged a target placed 230mm from the lens (sample images are shown in Figure 3) which is the minimum distance where the spherical aberration is negligible, so we can use MTF values calculated/measured for a distant object.

We grabbed 30 images, imported them into Matlab, and measured graylevels of a pixel in the white area and in the black area. The signal is given by the difference between the graylevels, as described by eq. 2.

$$signal = white\ level - black\ level \quad (2)$$

The noise is measured by evaluating the standard deviation of the pixel graylevel over the 30 images, i.e. temporal rather than spatial noise was measured, in order to null the effect of any spatial non-uniformity in the target illumination. Thirty images were used since this is the minimum number of samples needed to obtain a good estimate of the standard deviation[16]. To improve the accuracy of results, the signal and noise were then averaged over many pixels (a few hundred) in each of the areas (white and black).

In order to compare the measured noise to simulation, we want to obtain only the shot noise associated with the signal. Therefore, we need to subtract the noise in the black area (black squares in Figure 3 images), which is a result of shot noise from spurious diffraction order photons and readout noise associated with the camera electronics[17]. This is done according to eq. 3, where $\sigma$ stands for standard deviation, $\sigma_w$ is the noise in the white area of the image, $\sigma_b$ is the noise in the black area, and $\sigma_s$ the shot noise associated with the signal, as defined by eq. 2. The measured SNR is then given by eq. 4.

$$\sigma_s^2 = \sigma_w^2 - \sigma_b^2 \quad (3)$$
$$SNR = signal/\sigma_s \quad (4)$$

It should be noted that when evaluating an actual metalens design, the contribution of spurious diffraction orders to the noise should be considered, since they reduce the SNR by adding shot noise resulting from the background illumination, but not contributing to the signal (see section 4 of Supporting Information for guidelines on how to do this). However, for our current purpose of validating the theoretical model, it is better to measure and subtract this noise, since it is difficult to quantify it theoretically.

The simulated low-frequency SNR is calculated based on the number of photons reaching a camera pixel, assuming a shot-noise limited system (this is generally the case for practical modern systems operating in good lighting conditions). The SNR is therefore $\sqrt{N}$ where $N$ is the number of photoelectrons. The radiometric formulas for calculating the number of photoelectrons from the absolute power measured by the detector, the relative spectral distribution measured by the spectrometer, the spectral efficiency of the metalens, the spectral quantum efficiency of the cam-era, and the parameters of the optical relay system, are described in section 2 of the Supporting Information. The simulated SNR is then multiplied by the area under the simulated MTF of the metalens, per eq. 1, to obtain the simulated ASNR.

## Results

In Figure 2(a) the theoretical ASNR is shown, as a function of spectral range and aperture (the aperture is represented by the *F#*, defined as $F\# \equiv f/D$, where $f$ is the lens focal length and $D$ is the aperture diameter[18]). This simplified theoretical analysis assumes a Gaussian shaped spectrum, and wavelength independent efficiency of the metalens and camera (the spectral range was defined as $2\sqrt{2}\sigma$, where $\sigma$ is the standard deviation of the Gaussian distribution).

In Figure 2(b) we show the theoretical ASNR at an aperture of F/2.5 (a slice through the 2D graph of Figure 2(a), blue), alongside with a more accurate simulated ASNR (red) which takes into account our exact system parameters (measured spectral radiance, metalens and camera spectral efficiency). This is compared with the measured ASNR (yellow), extracted from experimentally captured images produced by our metalens. The spectral range of the band-pass filters was also defined as $2\sqrt{2}\sigma$, where $\sigma$ is the standard deviation of the distributions shown in Figure S1.

The results show that the optimum overall performance is obtained at a spectral range of approximately 50nm (this is independent of the absolute illumination level, which will change the absolute ASNR values, but not its spectral shape). The vertical offset between the simulated and measured results most likely results from a slight error in calculating the illumination level reaching the camera, which can be caused by inaccuracy in various parameters, e.g. the apertures of the imaging and illumination measurement systems.

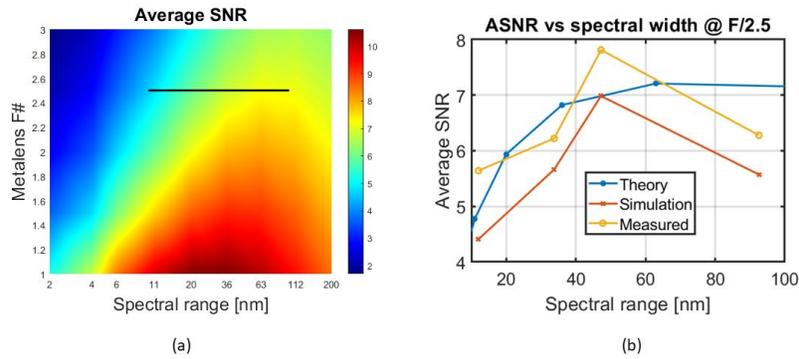

Figure 2. (a) ASNR vs. F# and spectral range. The black line marks the region where the 'Theory' results shown in 'b' are taken from. (b) ASNRs at F/2.5 –Theoretical, simulated and measured results.

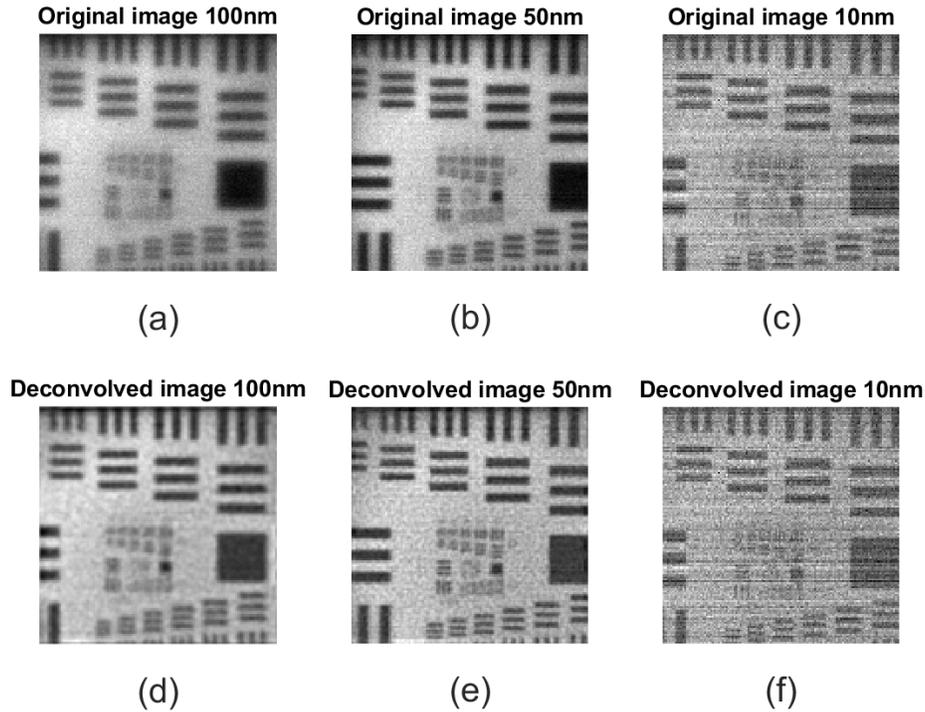

Figure 3. (a-c) Measured images at spectral ranges of 100nm, 50nm, and 10nm, respectively (d-e) Same as the above, following Weiner deconvolution.

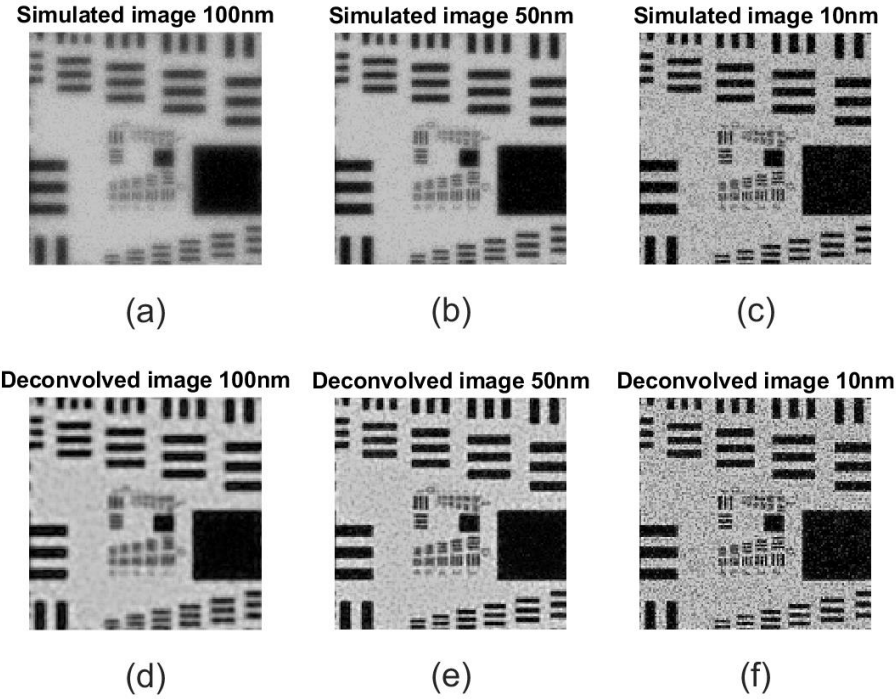

Figure 4. (a-c) Simulated images at spectral ranges of 100nm, 50nm, and 10nm, respectively (d-e) Same as the above, following Weiner deconvolution.

In Figure 3 the images of a resolution target taken using the metalens coupled to a CMOS camera (Thorlabs), at different spectral ranges, are shown. These images were purposely taken at low absolute radiance level and short camera exposure time ($25\ W/(m^2 \cdot nm)$ and $1.23\ ms$ respectively). While, as previously mentioned, the absolute illumination does not affect the optimal spectral range, for qualitative demonstration it is better to operate at low light level, in order to visually distinguish between different SNRs.

Looking at the raw images (Figure 3, a-c) one can see that the 100nm spectral range image is the blurriest, but less noisy. The 10nm image is sharpest, but noisier, and the 50nm image provides a convenient compromise between the two. Following Weiner deconvolution[10] (Figure 3, d-f), the resolution of the blurry images is improved, at the expense of added noise. The 50nm spectral range image still gives the best quality, as a compromise between resolution and noise. As a result of the low light level used, camera readout noise is visible in the 10nm bandwidth images, but this noise is subtracted out of the calculated ASNR (eq. 3), so it does not impact the 'measured' graph in fig 2b.

In order to facilitate the use of our ASNR metric in the design stage of a metalens, it is important to find out if our simulation can also provide qualitative image rendering that will give an indication of the expected image quality for a given design. To this end we used a high-quality image of a resolution target similar to the one used in the measurement. We then applied to it the simulated MTF (from Zemax) and SNR (the square-root of the number of photoelectrons, calculated based on the spectral radiance and exposure time mentioned above, integrated over the different spectral ranges). The result is shown in Figure 4.

The simulated images of Figure 4 resemble the measured images (Figure 3), with two main differences. The first difference is the low-frequency contrast, i.e. how black the large black areas are (known as 'veiling glare'[19]). This is because in our simulation we did not account for the light transmitted to other diffraction orders. This can, of course, be added to the simulation artificially, if the level of veiling glare is known from simulation or measurement – see more on this in section 4 of the Supplementary Information. Conversely, following the deconvolution process, the contrast of the real images can be enhanced by subtracting the black level from them (and then multiplying by an appropriate gain constant to raise the white level back up – this will of course increase the noise proportionally, so the SNR remains constant). An additional difference between the real and simulated images is visible mostly in the 10nm spectral width (Figure 3c,f vs. Figure4c,f). In the real images camera readout noise is visible, in the form of horizontal lines, and it is of course absent in the simulated images. This is not a serious impediment, since in most real-world scenarios one do not operate at such low light level/short exposure times.

## Conclusions

We have experimentally validated a theoretical method for evaluating the overall performance of a metalens system. The approach can work equally well for diffractive lens system. This method can be used to optimize operating spectral range of a chromatic or achromatic metalens design. It can also be used to compare performance of different designs, such as equivalent achromatic vs. chromatic designs. Examples of such applications will be shown in a follow-on paper.

## AUTHOR INFORMATION

### Corresponding Author

* Uriel Levy. Email: ulevy@mail.huji.ac.il

### Author Contributions

The manuscript was written through contributions of all authors. / All authors have given approval to the final version of the manuscript.

### Funding Sources

Ministry of Science and Technology

# How good is your metalens?
# Supporting Information

Setup hardware

The hardware used in the measurement setup shown in fig. 1 of the paper is as follows:

a. Light source – Thorlabs SLS201L (with fiber adapter removed)
b. Band pass filters – Thorlabs FB800-10, Thorlabs FB800-40, Salvo Technologies 2020OFS-825 NIR Bandpass Filter – 825nm FWHM 52nm, Salvo Technologies 102386944 NIR Bandpass Filter – 830nm FWHM 125nm.
c. Lightbox – An obsolete item. It is a Styrofoam lined box, with an illumination entrance port on the side, and an opal glass front end (an economical alternative to an integrating sphere, when high uniformity is not required).
d. Resolution target – Thorlabs R3L3S1P - Positive 1951 USAF Test Target, 3" x 3".
e. Metalens and iris – custom items, manufactured as part of our research by the nano-facilities of DTU and HUJI.
f. Imaging camera – Thorlabs DCC1545M. The reason we used this camera is that it has a removable C-mount adapter, that we could machine to mount our metalens. In addition, the camera window is near the detector plane, so it can accommodate the short BFL (back focal length) of our metalens.
g. Catalog lenses – Thorlabs. L1 – LA1433-B (f=150mm), L2 - LA1131 (f=50mm), L3 - LA1509-B (f=100mm).
h. Adjustable iris and fixed iris – Thorlabs SM1D12D
i. Removeable fold mirror – Thorlabs DFM1/M-P01
j. Alignment camera – FLIR BFS-U3-13Y3M
k. Optical Fiber - Thorlabs M28L01 - Ø400 µm, 0.39 NA
l. Spectrometer – Ocean Insight FLAME-T-XR1-ES
m. Detector – MKS-Ophir StarLite laser power meter and PD300R-UV sensor.

Radiometric calculation

In Figure 1 lenses L1 and L2 construct a relay optics that image the target plane onto the optical fiber entrance plane. We measure the relative spectral power that enters the fiber, using the spectrometer. In addition, by detaching the fiber output from the spectrometer, and attaching it to a power meter, we measure the absolute power entering the fiber.

In order to obtain the absolute spectral irradiance, we first normalize the relative spectral power ($E_{\lambda,rel}$) by dividing it by the area under the graph, so that the area under the normalized spectrum ($E_{\lambda,norm}$) is 1, as shown in eq. 1.

$$E_{\lambda,norm} = \frac{E_{\lambda,rel}}{\int E_{\lambda,rel}\, d\lambda} \tag{1}$$

To obtain the absolute spectral irradiance we must multiply $E_{\lambda,rel}$ by the total irradiance, obtained from the detector measurement. The detector measures power, and the irradiance is obtained by dividing the power by the fiber area (The fiber core diameter is 400µm. All the light that comes out of the relay optics enters the fiber, since its NA is larger than that of the relay optics, and all the light coming out of the fiber reaches the detector. We neglected losses originated from Fresnel reflections (~3.4% at the input and output of the fiber):

$$E_{ophir} = \frac{P_{ophir}}{A_{fiber}} \quad , \quad A_{fiber} = \pi(200 \times 10^{-6})^2 \tag{2}$$

Since the Ophir detector is calibrated for a specific wavelength (800 nm in our case), we introduce a correction factor when measuring a broadband source. The correction factor is calculated as follows:

$$corr\_factor = \int E_{\lambda,norm} \times \frac{R_{\lambda,ophir@800}}{R_{\lambda,ophir}} d\lambda \tag{3}$$

The spectral responsivity $R_{\lambda,ophir}$ used in the formula can be found at (we used the 'filter out' state):

https://www.ophiropt.com/laser--measurement/sites/default/files/PD300-UV_PD300-UV-193_PD300-IR_PD300-IRG_1.pdf

The accurate irradiance is now given by:

$$E_{ophir\_accurate} = corr\_factor \times E_{ophir} \tag{4}$$

The absolute spectral irradiance (in W/(m²nm)) at the entrance to the fiber is therefore given by:

$$E_\lambda = E_{\lambda,norm} \times E_{ophir\_accurate} \tag{5}$$

In order to find the spectral radiance $L_{\lambda,target}$ (or more accurately emittance) of the target, we must transfer the irradiance at the fiber plane back to the target plane. Since the target is approximately Lambertian we have[1]:

$$E_\lambda = \pi L_{\lambda,target} \times T_{lenses} \times (\frac{D}{2f})^2 \Rightarrow L_{\lambda,target} = \frac{E_\lambda}{\pi T_{lenses}} \times (\frac{2f}{D})^2 \tag{6}$$

$T_{lenses}$ is the transmission of the Thorlabs lenses used in the relay optics, and was calculated based on the Thorlabs coating data. D is the aperture of the fixed iris placed between the relay lenses (both operating at infinite conjugate) and $f$ is the focal length of L2.

Note: In our setup the radiance meter observes the resolution target at an angle. However, this does not affect the measured radiance, as a result of a law of radiometry that states that the radiance of a Lambertian source is independent of viewing angle[2].

Using the same formula, we can now calculate the irradiance at the metalens image plane:

$$E_{\lambda,cam} = \pi L_{\lambda,target} \times \tau_1(\lambda) \times (NA)^2 \tag{7}$$

Where $\tau_1(\lambda)$ is the metalens efficiency (fraction of light transmitted to first order of diffraction) as a function of wavelength. $NA$ is the metalens numerical aperture (in our case 0.2). The $NA$ is a function of wavelength, because of the strong chromatic aberration. It was assumed to be constant when computing the integral of eq. 8, since the variations are small. However, we did use different values for the different bandpass filters, based on the central wavelength (CWL) of the filter.

Finally, to calculate the number of photoelectrons N, we performed the following integral:

$$N = \frac{At}{hc} \int \eta(\lambda) \cdot \lambda \cdot E_{\lambda,cam} \, d\lambda \qquad (8)$$

Where $A$ is the pixel active area, $t$ is the exposure time, $\eta$ is the camera quantum efficiency, $h$ is Planck's constant, $c$ is the speed of light in vacuum, and $E_{\lambda,cam}$ is given by eq. 7.

The Thorlabs camera quantum efficiency can be found at:

https://www.thorlabs.com/newgrouppage9.cfm?objectgroup_id=4024&pn=DCC1545M

The SNR is now given by:

$$SNR = \sqrt{N} \qquad (9)$$

Supporting graphs

Measured relative spectral irradiance for the different bandpass filters is shown in Figure S1. Measured SNR and MTF, compared to simulation, are shown in figures S2, S3 respectively. Measured metalens spectral efficiency, measured using the method explained in [3], is shown in Figure S4. Camera and detector spectral efficiency were taken from manufacturer data (links given in section 2 above).

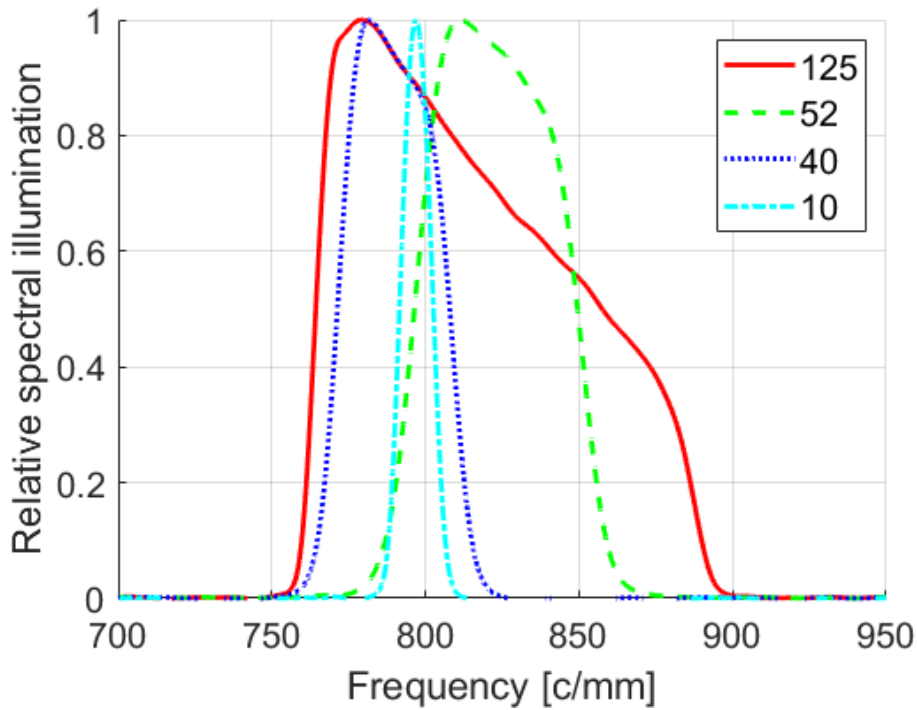

Figure S1: Relative spectral illumination, as output from spectrometer

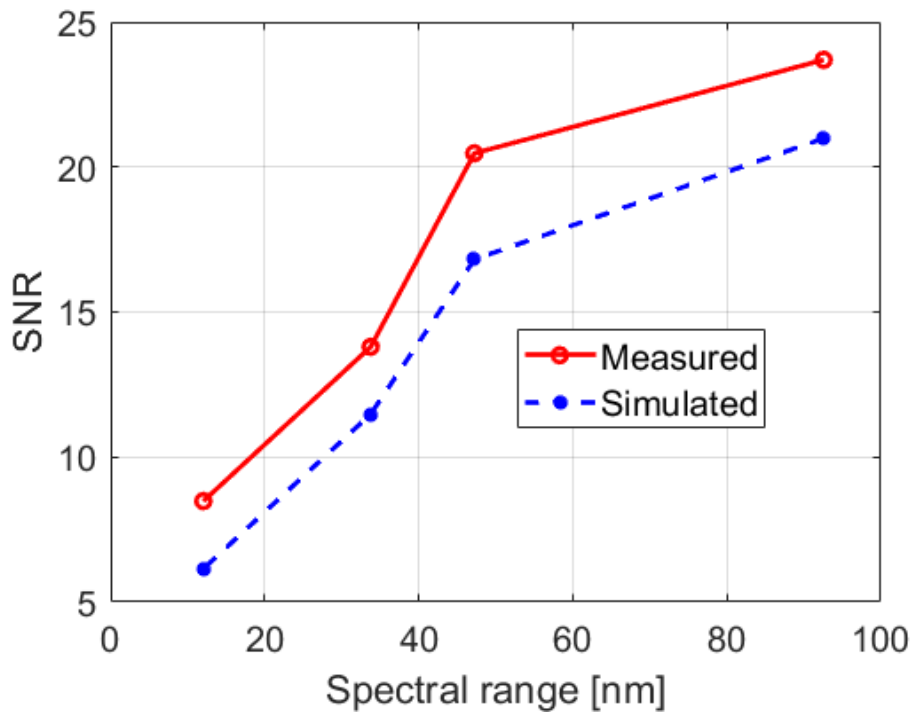

Figure S2: Measured vs. simulated SNR

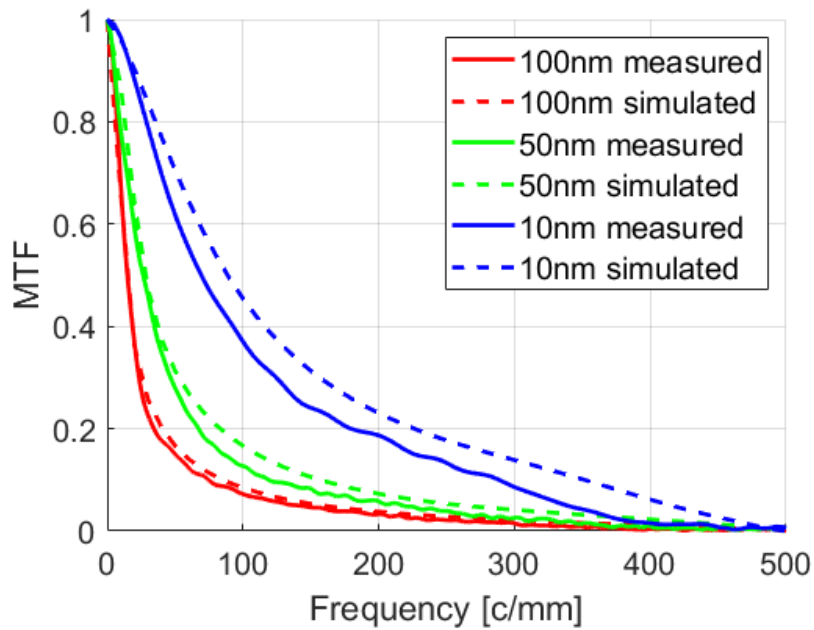

Figure S3: Simulated vs. measured MTF

From our experience the reason for the lower measured MTF results compared to simulation is a slight decenter of the mechanical iris with respect to the metalens optical axis, which introduces coma aberration[3].

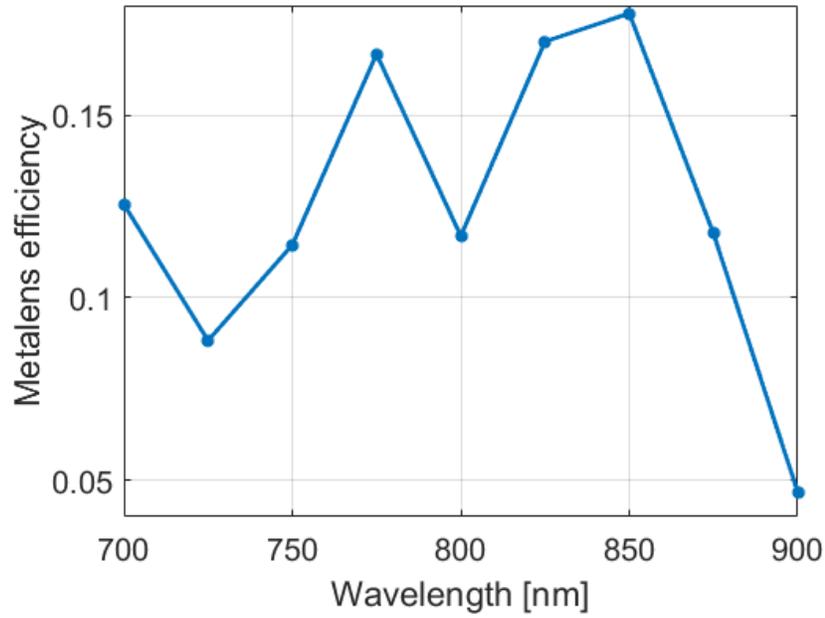

Figure S4: Measured metalens efficiency vs. wavelength

Effect of spurious diffraction orders

Veiling glare (VG) definition

Veiling glare of an image is generally defined as[4]:

$$VG \equiv \frac{black\ level}{white\ level} \qquad (10)$$

Where the black and white level are defined relative to the 'capped black' level of the camera, which is the average signal level output by the camera when no light is incident.

The modulation of a signal is defined as:

$$M \equiv \frac{white\ level - black\ level}{white\ level + black\ level} \qquad (11)$$

Combining eq. 10 and 11 we obtain eq. 12:

$$M \equiv \frac{1 - VG}{1 + VG} \qquad (12)$$

When the MTF is measured or calculated, it is generally normalized so that at zero frequency it is equal to 1. The parameter $M$ represents the absolute (un-normalized) MTF value at zero frequency, via eq. 12. So, we can multiply the MTF by this factor, to obtain the absolute MTF. Nevertheless, it is customary to separate the two effects, i.e. to use the normalized MTF and the VG as two separate metrics of system performance, the first representing the system 'resolution', whereas the second represents the system low-frequency 'contrast'.

Conventional refractive optical systems all have some level of VG, resulting from in-field light that undergoes multiple reflections from optical surfaces, and out-of-field light that is reflected from the internal mechanical housing - both types are called 'stray light'. If the stray light is distributed quasi-uniformly over the image, the effect is called veiling glare. Otherwise, it is called a 'ghost image' (if what is seen is a nearly focused image of the scene) or 'flare ' (if what is seen is a nearly focused image of the iris)[5].

In the case of a system that incorporates a metalens or diffractive lens we have an additional source of stray light – the spurious diffraction orders (i.e. light that goes to orders other than the design order, which is usually the first order).

Veiling glare measurement
During our SNR measurement described in the main paper, we measured the video white level, black level and capped black for the different spectral filters. Therefore, we can calculate the veiling glare as a function of spectral range, based on eq. 10. The result is shown in fig. S5. The variations in VG are mostly a result of the differences in CWL of the filters, and not their width, so the graph is a bit misleading in the sense that it does not really represent the dependence of VG on spectral width. However, for our purposes we can use the estimate $VG \approx 30\%$.

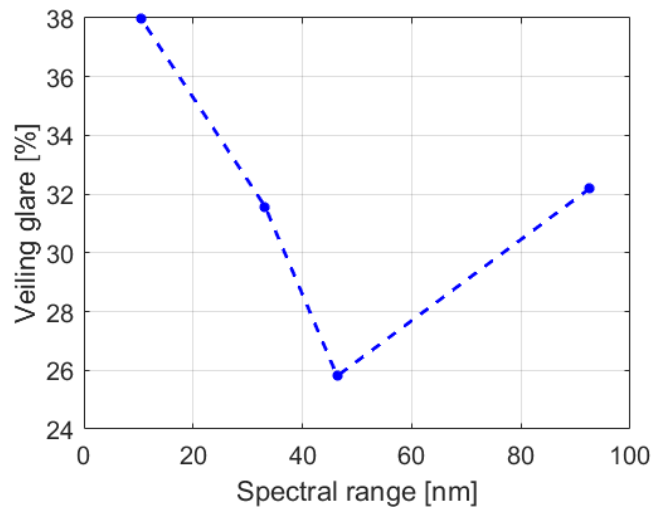

Figure S5: Veiling glare for the different filter widths

Analytic estimate of VG
Stray light analysis is usually performed numerically using commercial software packages – which may or may not support spurious diffractive order effects. It is difficult to analytically calculate the level of VG. However, an estimate for the case of a diffractive lens, where the VG is caused only by spurious diffraction orders, has been obtained by Buralli and Morris[6]. This estimate simply indicates that the zero-frequency modulation, which we will denote as $M_0$, is given by eq. 13, where $\tau_1$ is the average (over aperture and spectral range) diffractive lens efficiency, and $T$ is the average transmitted fraction of light (i.e. sum of efficiencies of all transmitted diffraction orders). The factor $T$ does not appear in [6] since they define $\tau_1$ relative to the transmitted light, while we defined it relative to the incident light. The normalized MTF (i.e. $MTF(0)=1$) can be multiplied by this factor to obtain absolute MTF values, that account for VG.

$$M_0 = \frac{\tau_1}{T} \tag{13}$$

In the case of our metalens, we measured the first order efficiency, and it is approximately 0.15 in the relevant spectral range (see fig. S4). In order to compare our measured VG results to the analytic estimate we must measure the total transmission (to all diffraction orders) of the metalens. This was done by illuminating the metalens with a collimated beam and placing a large area detector directly behind the lens. The ratio between this detector reading, and the detector reading when the metalens was removed but the mechanical aperture remained, gives us $T$.

We obtained a total transmission $T$ of about 55% (with some variation, depending on the spectral range). Substituting this and $\tau_1 = 0.15$ into eq. 13 we obtain that the expected zero frequency modulation is 0.27, and the expected VG (based on eq. 12) is 57%. This is higher than the measured result of about 30% (see previous section). We believe that the reason for this is that the analytic estimate is for the modulation of a low sinusoidal spatial frequency, so it implicitly assumes infinite extent of the illuminated area, and a duty cycle of 50% (i.e. half of the object plane is illuminated and half is dark). However, in our measurement setup the illuminated target covered a limited FOV of about ±8°. We therefore expect lower VG than obtained by the above analytic estimate. We have developed an analytic estimate for the case of a small object, but this is beyond the scope of the current paper.

### ASNR metric that includes VG effect

In our analysis, we subtracted both the signal and the noise of the dark region of the image. This was done since our purpose was to validate the theoretical model which is based on shot-noise. Since we knew the efficiency of our metalens for the first order of diffraction, but did not have accurate knowledge of the effects of the spurious orders, it was easier to make the comparison between theory and measurement using only the 'signal' from the first order of diffraction.

However, if the purpose is to characterize the overall performance of a metalens, perhaps in order to compare it to another, the veiling glare caused by stray light from spurious orders should be considered. To include the VG effects in the measurement all that needs to be done is **not** to subtract the black noise when calculating the signal noise, i.e. eq. 3 (of the main paper) will change to $\sigma_s^2 = \sigma_w^2$ , with no change to eq. 2, i.e. the signal is still the difference between the white and black levels, so the SNR of eq. 4 will decrease. This way we include also the camera readout noise, so we are characterizing the metalens + camera system. If we are interested in characterizing the metalens, without the camera, we can subtract the readout noise by subtracting the capped black noise, i.e. the noise when no light reaches the camera.

To include the VG effects in simulation of performance, we must calculate the SNR differently. We can still use eq. 8 to calculate the number of signal photoelectrons, but eq. 9 is no longer valid, since the noise is equal to the square root of the total no. of photoelectrons, including those created by the stray light. To calculate the total noise, we must replace $\tau_1$ in eq. 7 with $T$, and substitute into equation 8, to obtain $S_N$, the signal including background. Instead of equation 9, we then have:

$$SNR = S/\sqrt{S_N} \qquad (14)$$

This can be substituted into eq. 1 of the main paper to obtain the simulated ASNR (using the normalized MTF, without the factor of eq. 13).

Note that we have not accounted for FOV in the current definition of our ASNR metric, since the images shown cover a FOV of about ±8°, and our efficiency does not change much over this range[3]. Our MTF actually does drop significantly over this range (as a result of lateral chromatic aberration)[3], but in this paper we chose to refer only to on-axis performance, since our purpose was only validate the ASNR metric. If one wants to compare overall performance of wide-FOV metalenses, and account also for performance variation over the FOV, one can redefine the ASNR as an average (or a weighted average, if one wants for example to give more weight to the on-axis performance) over the ASNR at several locations in the FOV.

Camera characterization

Before we could use our camera to measure SNR at different spectral ranges, we had to characterize it to see if it fulfills two requirements: (a) Linearity (b) Shot noise limited performance. If this is not the case we cannot expect to obtain an SNR that follows the $\sqrt{N}$ formula (if the camera is not linear, the signal will not follow $N$, the no. of photoelectrons, and if the camera is not shot-noise limited the noise will not follow $\sqrt{N}$).

We used the same experimental setup as that described in the main paper, but instead of using the same illumination level, and exchanging spectral filters, we used a single spectral filter (we chose a 40nm width filter, but it is not important) and varied the illumination level using the iris attached to the light source. The camera exposure time was chosen so that at the maximum illumination level and minimum camera gain, we have a white level of about 200, i.e. as high as possible, without saturating (the maximum graylevel in our camera is 255, i.e. 8-bit output). We then reduced the illumination measured by the detector by a factor of 2 each time, and measured the absolute signal level (Ophir detector reading), the camera signal level (mean white graylevel value) and camera noise (standard deviation of pixel in white area). In order to avoid being limited by quantization noise (caused by conversion of the analog pixel output voltage to discrete graylevel values), we increased the camera gain at the low illumination levels, to bring the signal back up to graylevel of about 200. We than divided by the gain during the calculation, to obtain the final signal and noise levels.

In fig S7 we can see the raw signal transfer function (STF, i.e. camera output vs. illumination input) and photon transfer curve (PTC, this is the term used for the camera noise as a function of illumination input)[7]. In fig. S8 we show the PTC in log-log scale, and compare it to the textbook case of a readout noise/shot noise limited system. One can clearly see the approximate 0.5 slope at the higher signals – indicating a shot noise limited system, and the flattening out at low signal – as a result of dark noise. In fig S9 we plot the STF and PTC on log-log scale, but this time with the capped black subtracted, from both signal and noise respectively. From these graphs we can obtain the slope of the linear fit. We obtain a slope near 1 for the STF, indicating linearity, and a slop near 0.5 for the PTC, indicating that we are nearly shot noise limited (The slope of the PTC with black noise subtracted of fig. S9b is a bit higher that obtained without the black noise subtracted in fig S8a. This is because in fig S8a we fit only the higher illumination points).

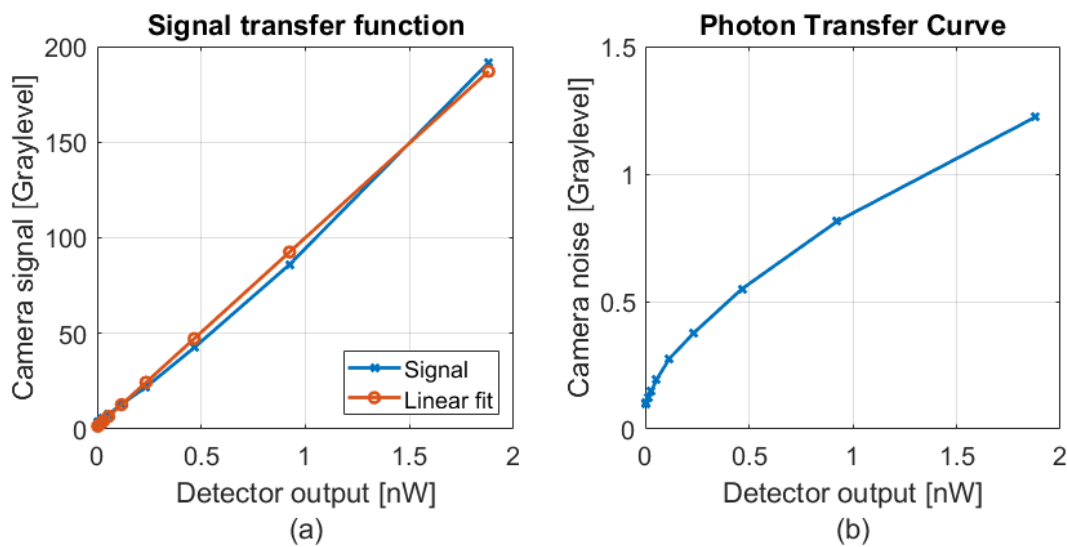

Figure S6: Camera transfer curves – linear scale without subtraction of capped black

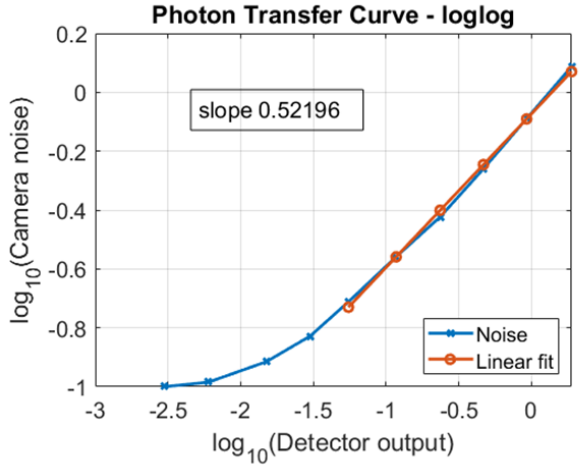
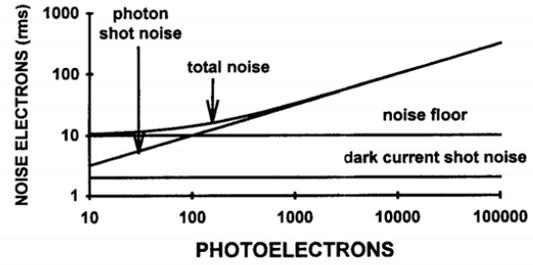

(a)

(b)

Figure S7: Camera transfer curves – log-log scale without subtraction of capped black

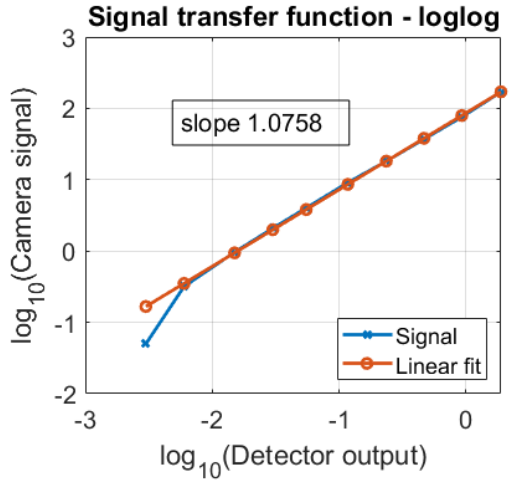
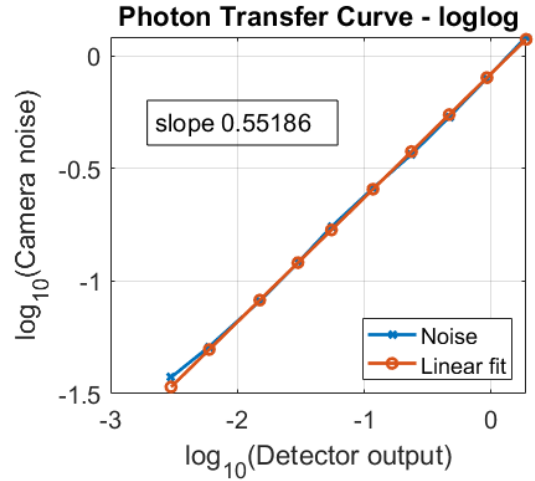

(a)

(b)

Figure S8: Camera transfer curves – log-log scale with subtraction of capped black

MTF calculation

The MTF used in the calculation of the theoretical and simulated ASNR shown in fig.1(a,b) of the main paper is based on a Fraunhofer approximation. The Fraunhofer MTF at each wavelength was computed using Zemax optical design software and a weighted sum of these MTFs was performed to obtain to overall performance (in the general case of a non-symmetric PSF one must sum OTFs, optical transfer functions, and not MTFs, since the MTF is the absolute value of the OTF, so the phase information is lost. However, for our on-axis analysis the PSF is symmetrical, so the MTF and OTF are the same).

In the Supplementary Information of [3] the validity of the Fraunhofer approximation for the calculation of our chromatic defocus PSFs was checked. It was found that for a blur spot radius smaller than 0.25mm the approximation is valid. Using the expression for chromatic aberration of a diffractive lens given in [8] we can find the maximum spectral width for which the approximation is valid:

$$x_{2max} \approx \Delta f \cdot NA = f \frac{\Delta \lambda}{\lambda} \cdot NA = 3.36 \frac{\Delta \lambda}{800} \cdot 0.2 = 0.25 \Rightarrow \Delta \lambda \approx 300 nm \qquad (15)$$

The $\Delta\lambda$ of the eq. 15 is the wavelength shift from nominal, so seemingly the total spectral width can be up to 600nm. But since we defined our spectral width as $2\sqrt{2}\sigma$, a spectral width of 600nm will contain wavelengths that are shifted more than 300nm from the nominal. In any case, the approximation should hold well for our experimental spectral widths of up to 100nm. In fig. S9 we show the ASNR graphs up to broader spectral widths. When the spectral width exceeds about 300nm (depending on the aperture) the results begin to diverge.

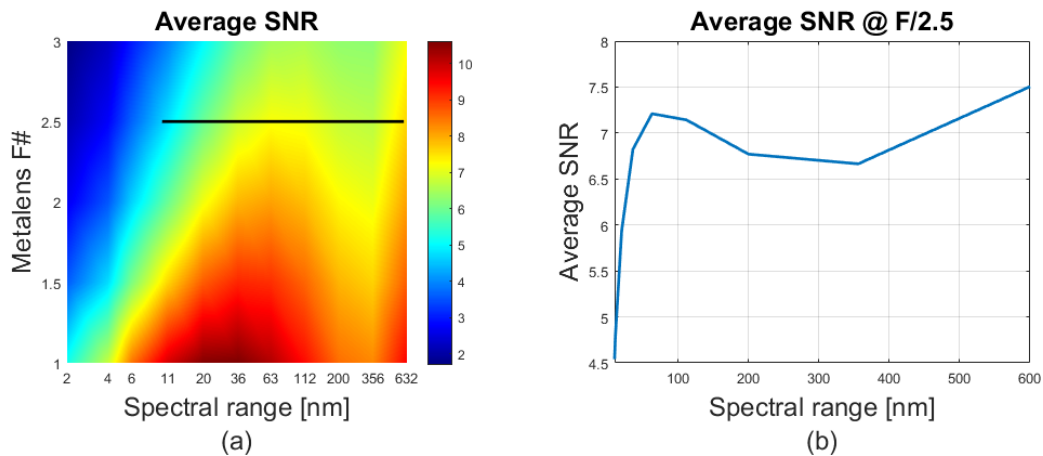

Figure S9: Fraunhofer based theoretical ASNRs – extended spectral range